\documentclass[rnote]{aa} 

\usepackage{amsmath}
\usepackage{graphicx}
\usepackage{longtable}
\usepackage{natbib}
\usepackage{txfonts}
\usepackage{amsfonts}
\usepackage{amssymb}
\usepackage{url}
\usepackage{subfigure}

\bibpunct{(}{)}{;}{a}{}{,}

\begin{document}

\title{The masses of the neutron and donor star in the high-mass X-ray binary IGR J18027-2016 \thanks{Based on observations carried out at the European Southern Observatory under programme ID 085.D-0539(A)}}

\author{A.B. Mason \inst{1}
\and A.J. Norton   \inst{1}
\and J.S. Clark  \inst{1}
\and I. Negueruela \inst{2}
\and P. Roche \inst{1,3,4}}

\institute{Department of Physics \& Astronomy, The Open University, Milton Keynes MK7 6AA, UK \and
 Departamento de F\'{\i}sica, Ingenier\'{\i}a de Sistemas y
  Teor\'{\i}a de la Se\~{n}al, Universidad de Alicante, Apdo. 99,
  E03080 Alicante, Spain \and
  School of Physics \& Astronomy, Cardiff University, The Parade, Cardiff, CF24 3AA, UK. \and
  Division of Earth, Space \& Environment, University of Glamorgan, Pontypridd, CF37 1DL, UK.}

\date{Received 02 June 2011 / Accepted 02 July 2011}

\abstract{}
{We report near-infrared observations of the supergiant donor to the eclipsing high mass X-ray binary pulsar \object{IGR J18027-2016}. We aim to determine its spectral type and measure its radial velocity curve and hence determine the stellar masses of the components.}
{ESO/VLT observations of the donor utilising the NIR spectrograph ISAAC were obtained in the H and K bands. The multi-epoch H band spectra were cross-correlated with RV templates in order to determine a radial solution for the system.}
{The spectral type of the donor was confirmed as B0-1 I. The radial velocity curve constructed has a semi-amplitude of $23.8 \pm 3.1$~km~s$^{-1}$. Combined with other measured system parameters, a dynamically determined neutron star mass of 1.4 $\pm$ 0.2 - 1.6 $\pm$ 0.3 ~M$_{\odot}$ is found. The mass range of the B0-B1 I donor was 18.6 $\pm$ 0.8 - 21.8 $\pm$  2.4 ~M$_{\odot}$. These lower and upper limits were obtained under the assumption that the system is viewed edge-on (i = 90$^\circ$ with $\beta$ = 0.89) for the lower limit and the donor fills its Roche lobe ($\beta = 1$ with i = 73.1$^\circ$) for the upper limit respectively.} 
{}

\keywords{binaries:eclipsing - binaries:general - X-rays:binaries - stars:individual:IGR J18027-2016}

\authorrunning{A.B. Mason et al}
\titlerunning{Mass of the NS and donor in HMXB IGR J18027-2016} 

\maketitle

\begin{table*}
\caption{The phase, radial velocity and telluric standard for each \object{IGR~J18027-2016} spectrum.}
\centering
\begin{tabular} {ccccc}
\hline
Mid-point of Observations (UT) & MJD & Phase & Radial velocity / km s$^{-1}$ & Telluric Std \\
\hline
 2010 July 13.161 & 55390.16060 & 0.073 & 51.8 $\pm$ 9.9  & Hip 088201  \\
 2010 July 09.017 & 55386.01693 & 0.166 & 76.9 $\pm$ 9.9 & Hip 089744  \\
 2010 April 22.336 & 55308.34560 & 0.169 & 68.4 $\pm$ 9.9 & RV std  \\
 2010 June 07.210 & 55354.21010 & 0.205 &  85.2 $\pm$ 9.9 & Hip 092931  \\
 2010 Aug 10.197 & 55418.19702 & 0.208 & 79.9 $\pm$ 9.9 & Hip 092322  \\
 2010 Sept 06.995 & 55445.99528 & 0.291 & 76.0 $\pm$ 9.9 & Hip 094859  \\
 2010 June 12.252 & 55359.25220 & 0.309 & 69.2 $\pm$ 9.9 & RV std  \\
 2010 June 22.074 & 55369.07022 & 0.457 & 64.8 $\pm$ 9.9 & RV std  \\
 2010 Aug 25.171 & 55433.17163 & 0.485 & 47.7 $\pm$ 9.9 & Hip 100170  \\
 2010 Aug 16.111 & 55424.11101 & 0.502 & 48.6 $\pm$ 9.9 & Hip 092470  \\
 2010 Sept 08.080 & 55447.07987 & 0.529 & 33.4 $\pm$ 9.9 & Hip 090804  \\
 2010 July 29.211 & 55406.21077 & 0.585 & 42.6 $\pm$ 9.9 & Hip 089384  \\
 2010 June 09.095 & 55356.09537 & 0.618 & 37.8 $\pm$ 9.9 & Hip 083535  \\
 2010 May 26.392 & 55342.39190 & 0.619 & 22.7 $\pm$ 9.9 & RV std   \\
 2010 Aug 08.104 & 55416.10351 & 0.750 & 19.9 $\pm$ 9.9 & Hip 101552  \\
 2010 Aug 31.054 & 55439.05354 & 0.772 & 24.2 $\pm$ 9.9 & Hip 094378  \\
 2010 June 19.167 & 55366.16684 & 0.822 & 56.4 $\pm$ 9.9 & HD 169101  \\
 2010 July 08.191 & 55385.19053 & 0.985 & 56.3 $\pm$ 9.9 & Hip 090978  
\end{tabular}
\end{table*}

\section{Introduction}
The precise nature of the fundamental properties of neutron star (NS) matter has been an on-going field of research for many decades. 
The physical properties of matter that exist in the extreme densities and pressures found within a NS can be determined by the NS equation of state (EoS). At this juncture however there exist over 100 proposed EoS \citep{kaper06}. Only one can be physically valid. Observational data can assist in reducing this number by eliminating contending EoS that place unrealistic constraints on the mass range of observed NSs. \\* Many EoS predicting the presence of exotic hadronic matter have been ruled out by the recent mass determination of 1.97 $\pm$ 0.04 M$_{\odot}$ for the binary millisecond pulsar (MSP) \object{PSR J1614-2230} \citep{demorest10}. In the case of \object{PSR J1614-2230} it is believed that the over-massive NS in this system was created by the accretion of matter whilst being spun-up to become a MSP. Another possible means of high mass NS production may occur where massive progenitors have yielded a high pre-SN core mass. High Mass X-ray Binary systems (HMXBs) are potential testing grounds for this idea as in the majority of cases the progenitor of the NS will be more massive than the donor star we can observe. \\* The only means of performing unambiguous NS mass determinations in X-ray pulsar binaries is by observing eclipsing binary systems in which we can constrain the inclination angle. There are presently 10 such systems known with 8 in which the NS mass has been calculated (e.g. \citet{mason11}; \citet{mason10}; \citet{quaintrell03}). In this paper we present an orbital solution from observations of the donor within the HMXB accretion driven pulsar IGR J18027-2016. The mass donor within this system is heavily obscured and reddened. Using near infrared spectroscopy of the donor star we have calculated an orbital solution to measure the mass of each of the components within the system.  \\
\object{IGR J18027-2016} was first detected by {\it INTEGRAL} in 2003, \citep{rev04} and was spatially associated with the X-ray pulsar \object{SAX J18027.7-2017} which was discovered serendipitously during observations of the LMXB \object{GX9+1} and found to have a pulse period of 139.6 s  \citep{augello03}. \object{IGR J18027-2016} has a measured hydrogen column density, N$_{\rm H}$ = 9.1 $\pm$ 0.5 $\times$ 10$^{22}$ cm$^{-2}$ \citep{walter06} greatly in excess of the line of sight column density of N$_{\rm H}$ = 1.0 $\times$ 10$^{22}$ cm$^{-2}$. This is indicative of intrinsic absorption occurring within the system, with matter from the stellar wind of the donor forming a dense spherical shell around the NS as it is accreted \citep{hill05}.\\
{\it XMM-Newton} observations in 2004 narrowed the source position down (with an uncertainty of 4$^{\prime\prime}$) to the location RA(2000.0) = 18$^{h}$02$^{m}$42.0$^{s}$ and Dec = -20$^{\circ}$17$^{\prime}$18$^{\prime\prime}$ \citep{walter06}. Using this improved position \citet{masetti08} were able to identify the donor as \object{2MASS J18024194-2017172} (J = 12.7, H = 11.9, K = 11.5).   \\
From timing analysis \object{IGR J18027-2016} was found to have an orbital period of 4.5696 $\pm$ 0.0009 days \citep{hill05}. The orbit based upon the sinusoidal modulation of pulse arrival times was found to be approximately circular. Using a donor mass-radius relation together with approximations of the Roche lobe radius  \citet{hill05} proposed that the donor is a 09 - B1 supergiant. \\
Optical and NIR spectral analysis of the mass donor performed by \citet{chaty08} detected prominent Paschen and Brackett series hydrogen lines in addition to He {\sc i} and He {\sc ii} lines in emission. Combining their spectroscopy with SED modelling they proposed that the mass donor was an early B supergiant. Obtaining higher resolution spectra in the I and K bands \citet{torrejon10} narrowed the spectral type down to B1 Iab - B1 Ib. The short orbital period of \object{IGR J18027-206} of only 4.6 d eliminated the donor being a B1 Iab supergiant, they proposed a classification of B1 Ib for the mass donor lying at a distance of 12.4 kpc.

\section{Observations and Data Reduction}
Observations were conducted between 2010 May 26th and 2010 Sept. 08th using the NIR spectrograph ISAAC on the VLT in the SW MRes mode utilising a 0.8$^{\prime\prime}$ slit width. Although the mass donor in \object{IGR J18027-2016} is relatively faint (H = 11.91) we were able to obtain medium resolution (R~$\sim$~3000) and high S/N spectra in the H and K$_{s}$ bands. For the H band multi-epoch science and RV template exposures centred on 1.7 $\mu$m were obtained of 2400s and 200s respectively. For the K$_{s}$ band two science exposures centred on 2.06 and 2.15 $\mu$m were taken for a total of 2240s. Reduction was performed using the ISAAC pipeline with OH skylines used to wavelength calibrate the spectra. The resulting data has a count rate less than 10 000 ADU and thus no correction for detector non-linearity was necessary. The RV template observed was Hip 89262, with a known RV of -8 km s$^{-1}$, close in spectral type to the target (B0.5 Ia) and bright (H~$\sim$ 6.8).\\
We obtained a dataset containing 18 spectra. Telluric correction was employed to remove atmospheric features from each spectrum. Unfortunately, no telluric standards were obtained for the first four spectra in our dataset (see Table 1). In this case we employed the RV template (which had an airmass close to that of the target) as a telluric standard. In order not to contaminate our target spectrum we first removed any non-telluric spectral lines from the RV template before telluric correction was attempted. The continuum normalised H band spectra comprising our full dataset ordered by phase are shown in Fig. 1.

\begin{figure}[!htb]
    \includegraphics[width=8cm,angle=0.2] {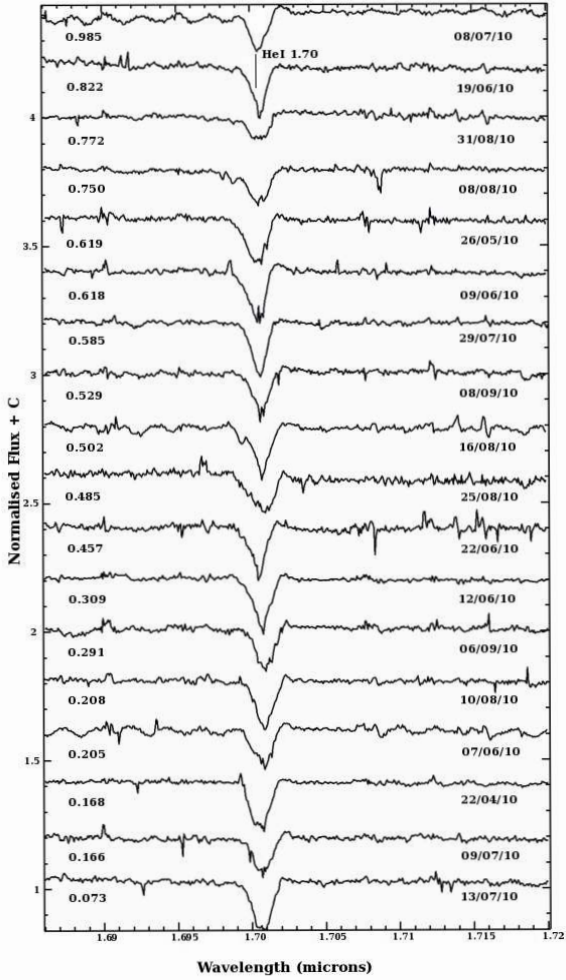}
    \begin{center}
    \caption{Continuum normalised H-band spectra centred on He {\sc i} 1.70 $\mu$m of \object{IGR~J18027-2016} in order of phase from bottom to top.}
    \end{center}
\end{figure}

\section{Data Analysis}
Using the standard IRAF\footnote{IRAF is distributed by the National Optical Astronomy Observatory, which is operated by the Association of Universities for Research in Astronomy, Inc., under cooperative agreement with the National Science Foundation.} routine {\it fxcor}, radial velocities were determined by cross-correlating the region around the He {\sc i} 1.70 $\mu$m absorption line in the 8 science spectra against the same region in the RV template spectra which were obtained shortly after each science exposure. The resulting radial velocities were then corrected to the solar system barycentre and are reported in Table 1.   
The eccentricity of the system is reported as e $\lesssim$ 0.2 \citep{augello03} and although no precise measurement of the eccentricity is reported, \citet{hill05} found sinusoidal modulation from an analysis of pulse arrival times that indicates the NS is in a circular orbit around the mass donor. We have thus fitted the radial velocities of the supergiant donor with a sinusoidal solution, using the ephemeris of \citet{hill05} which specifies the epoch of mid-eclipse as 
\begin{equation}
{\it T}(MJD) = 52168.26(4) + 4.5696(9){\it N}
\end{equation}
the uncertainties in brackets refer to the last decimal place quoted and {\it N} is the cycle number. At the epoch of our observations the accumulated uncertainty in phase is 0.11. We therefore fitted our data with two models, one in which the zero phase is fixed and the other in which we allowed the zero phase to vary as a free parameter. 

In our first model we fixed the zero phase and allowed the systemic velocity and RV semi-amplitude to vary as free parameters. Fitting a sinusoid to our data we found the radial velocity semi-amplitude K$_{\rm O}$ = 24.4 $\pm$ 3.2 km s$^{-1}$ and systemic velocity $\gamma$ = 51.7 $\pm$ 2.4 km s$^{-1}$. To obtain this solution the uncertainties in RV for each data point had to be scaled to $\pm$ 9.9 km s$^{-1}$ thus reducing chi-squared to unity. \\
Alternatively fitting a sinusoid to our data with three free parameters (allowing in this case the zero phase to vary) we found the radial velocity semi-amplitude K$_{\rm O}$ = 23.8 $\pm$ 3.1 km s$^{-1}$, systemic velocity $\gamma$ = 52.5 $\pm$ 2.4 km s$^{-1}$ and a phase shift of -0.03 $\pm$ 0.02. Here the uncertainties were scaled to $\pm$ 9.7 km s$^{-1}$. The best-fit phase offset found is within the accumulated phase uncertainty of the ephemeris. We prefer this fit and use the data from it for our subsequent calculations of the system parameters of \object{IGR J18027-2016}. Both fits are shown in Fig 2.

\begin{figure}
 \includegraphics[width=9cm]{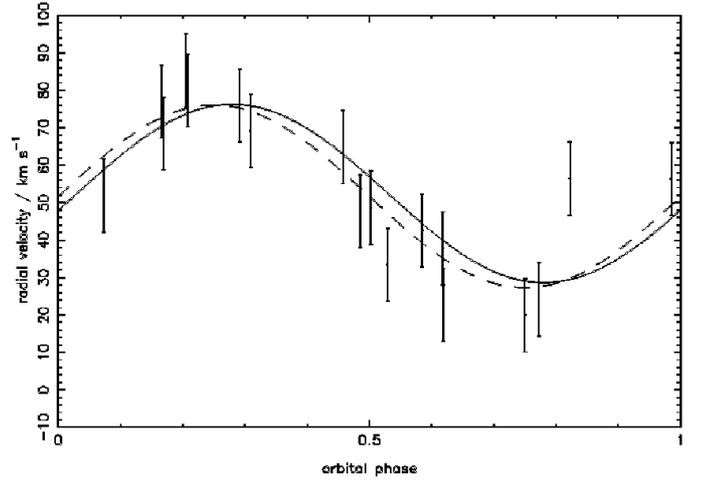}
    \begin{center}
    \caption{Radial velocity data for the supergiant mass donor in \object{IGR J18027-2016}. The solid line is the best fitting sinusoid with three free parameters, the dashed line is that with a fixed zero phase in line with the published ephemeris. The orbital phase is based upon the ephemeris of \citet{hill05}.}
    \label{rvcurve}
    \end{center}
\end{figure}

\begin{table}
\caption{System parameters for \object{IGR J18027-2016}.} 
 \label{results}
 \begin{tabular}{llll} \hline
Parameter & \multicolumn{2}{c}{Value} & Ref. \\ \hline
{\it Observed}     		&	&	& \\
$a_{\rm X} \sin i$ / lt sec	& \multicolumn{2}{c}{$68 \pm 1$}		& [1]\\
$P$ / d				& \multicolumn{2}{c}{$4.5696 \pm 0.0009$}	& [1]\\
$T_{90}$ / MJD		& \multicolumn{2}{c}{$52 168.26 \pm 0.04$}	        & [1]\\	
$e$				& \multicolumn{2}{c}{$\lesssim$ 0.2}		& [2]\\	
$\theta_{\rm e}$ / deg		& \multicolumn{2}{c}{$34.9 \pm 4.6$}		& [1]\\	

$K_{\rm O}$ / km s$^{-1}$	& \multicolumn{2}{c}{$23.8 \pm 3.1$} 		& [3]\\

{\it Assumed}     		&	&	& \\
\bf{$\Omega$}			& \multicolumn{2}{c}{= 1}		     \\

{\it Inferred} & & & \\

$K_{\rm X}$ / km s$^{-1}$	& \multicolumn{2}{c}{$324.4 \pm 4.7$}  		\\
$q$				& \multicolumn{2}{c}{$0.073 \pm 0.01$}  \\
$\beta$				& $1.000$ 	& $0.893 \pm 0.078$	 \\
$i$ / deg			& 73.1 $\pm ~6.3$& $90.0$                 \\
$M_{\rm X}$ / M$_{\odot}$ 	& 1.58 $\pm ~0.27$ & $1.36 ~\pm$ 0.21 	 \\
$M_{\rm O}$ / M$_{\odot}$ 	& 21.8 $\pm $ 2.4 & $18.6 \pm 0.8$ 	 \\
$a$ / R$_{\odot}$ & $33.1 ~\pm$ 1.2 & $31.4 \pm 0.5$  \\
$R_{\rm L}$ / R$_{\odot}$	& 19.8 $\pm$ 0.8  & $18.8 ~\pm$ 0.3   \\
$R_{\rm O}$ / R$_{\odot}$	& 19.8 $\pm$ 0.7  & $16.8 ~\pm$ 1.5           \\ \hline
\end{tabular}\\
$[1]$ \citet{hill05}; $[2]$ \citet{augello03}\\
$[3]$ this paper
\end{table}

\begin{figure*}[!htb]
    
    \begin{center}
    \subfigure  {\includegraphics[width=9cm] {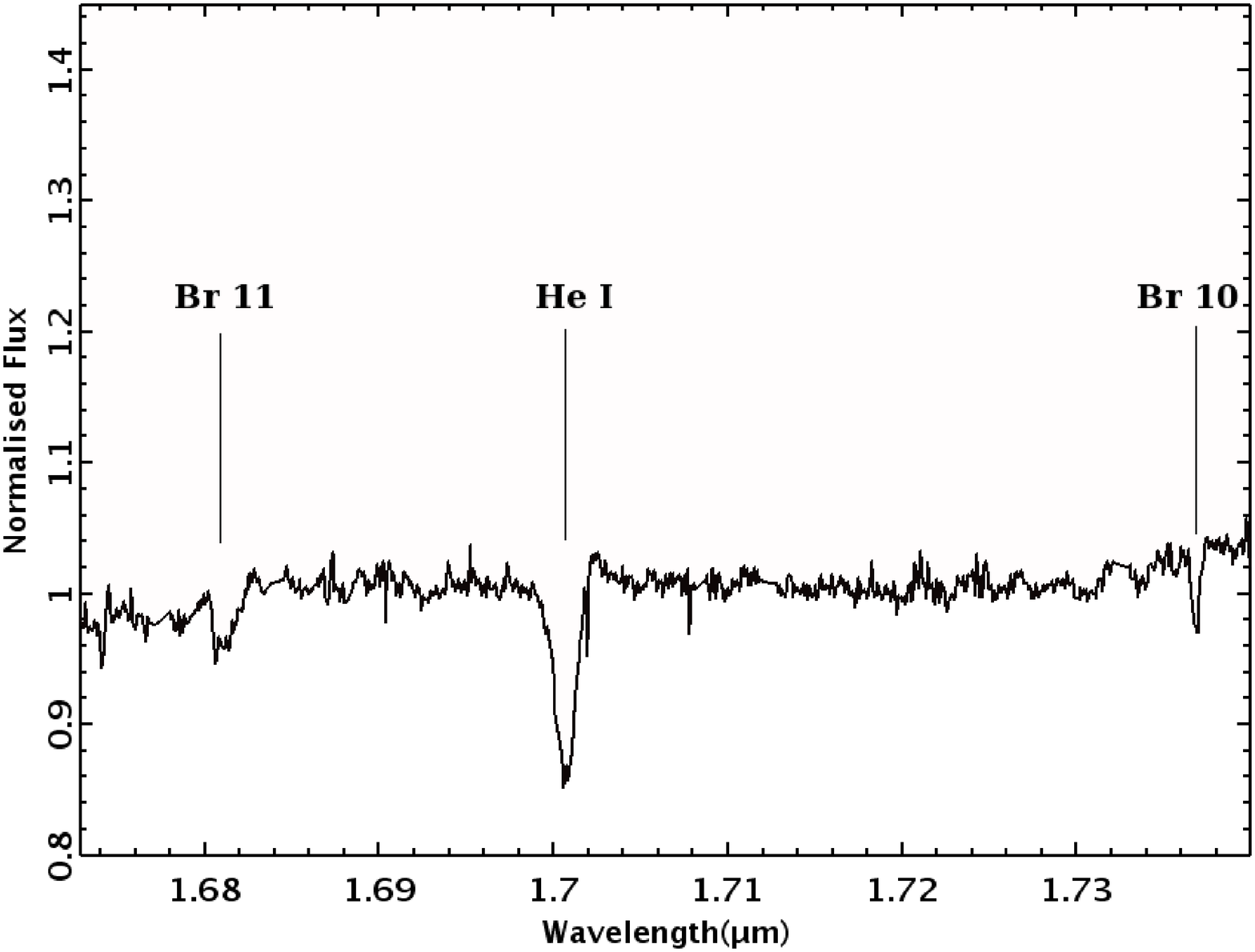}}
    \subfigure  {\includegraphics[width=9cm] {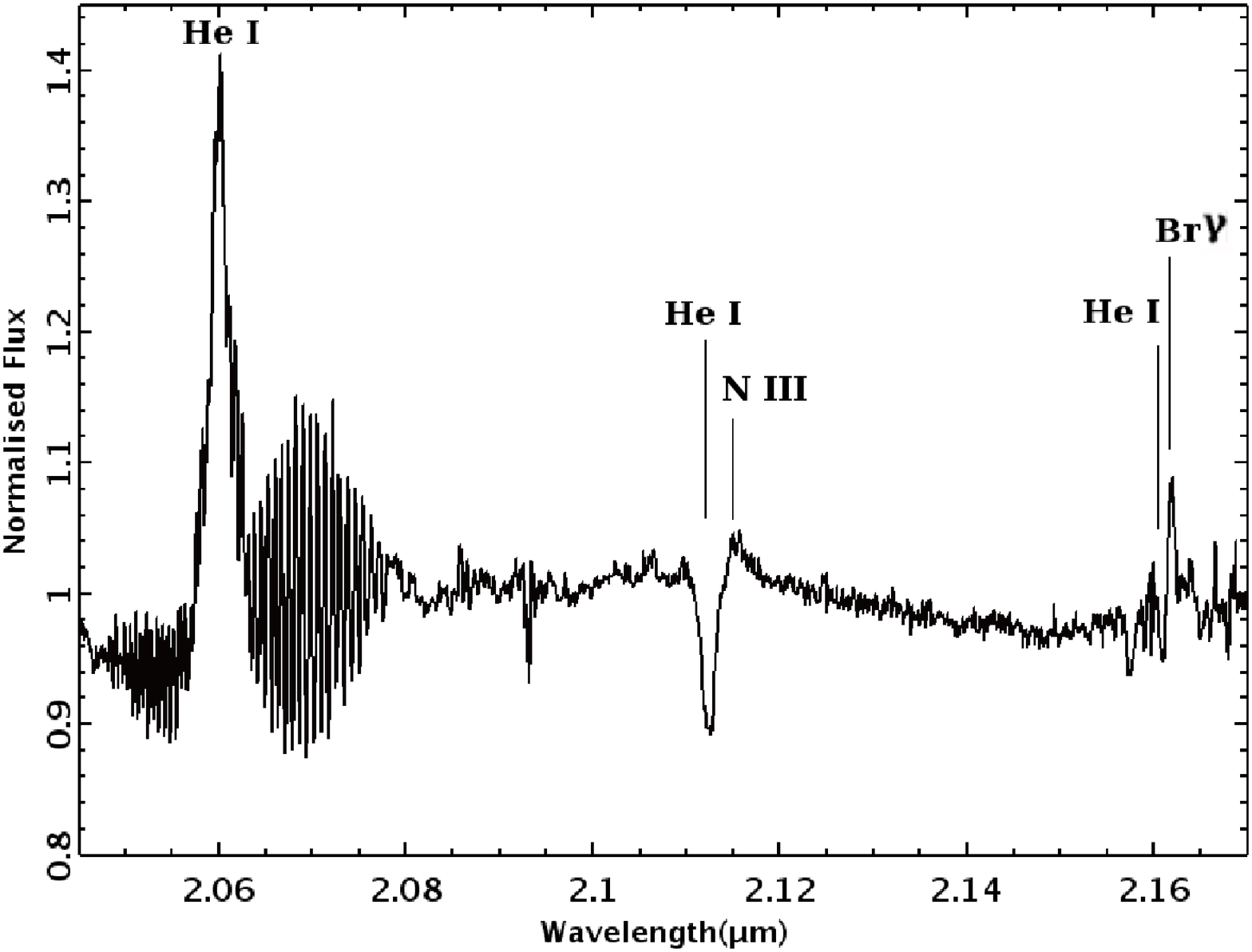}}
    \caption{{\it Left}: IGR J18027-2016 continuum normalised H band spectrum. {\it Right}: Continuum normalised K$_{s}$ band spectrum of IGR J18027-2016.}
    \end{center}
\end{figure*}

The projected semi-major axis of the neutron star's orbit from X-ray pulse timing delays was measured by \citet{hill05} as $a_{\rm x} \sin i = 68.0 \pm 1$~light seconds. From this, the semi-amplitude of the neutron star's radial velocity may be calculated using
\begin{equation}
    a_{\rm X} \sin i = \left(\frac{P}{2\pi} \right) K_{\rm X} 
\end{equation}
for a circular orbit to give $K_{\rm X} = 324.4 \pm 4.7$~km~s$^{-1}$.

To determine the component masses of the system we must first consider the mass ratio of the system $q$ which is equal to the ratio of the semi-amplitudes of the radial velocities for each star
\begin{equation}
     q = \frac{M_{\rm X}}{M_{\rm O}} = \frac{K_{\rm O}}{K_{\rm X}}
\end{equation}
where $M_{\rm X}$ and $M_{\rm O}$ are the masses of the neutron star and supergiant star respectively, and
$K_{\rm X}$ and $K_{\rm O}$ are the corresponding semi-amplitudes of their radial velocities. In addition, for circular orbits,
\begin{equation}
M_{\rm O} = \frac{{K_{\rm X}}^3 P}{2\pi G \sin^3 i}\left(1+q\right)^2
\end{equation}
and similarly
\begin{equation}
M_{\rm X} = \frac{{K_{\rm O}}^3 P}{2 \pi G \sin^3 i}\left(1+\frac{1}{q} \right)^2
\end{equation}
where $i$ is the inclination to the plane of the sky and $P$ is the orbital period.  The system inclination can be found from the geometric relation
\begin{equation}
     \sin i \approx \frac{\left[1 - \beta^2 \left(\frac{R_L}{a} \right)^2\right]^{1/2}}{\cos~\theta_{\rm e}}
\end{equation}
where $\theta_{\rm e}$ is the eclipse half-angle, $R_{\rm L}$ is the Roche lobe radius of the supergiant, $\beta$ is the ratio the supergiant's radius to that of its Roche lobe and $a$ is the separation between the centres of mass of the two stars. The Roche lobe radius may be approximated by
\begin{equation}
   \frac{R_{\rm L}}{a} \approx A + B \log q + C \log^2 q
\end{equation}
where the constants have been determined by \citet{rappaport84} as
\begin{equation}
A \approx 0.398 - 0.026\Omega^2 + 0.004\Omega^3
\end{equation}
\begin{equation}
B \approx - 0.264 + 0.052\Omega^2 - 0.015\Omega^3
\end{equation}
\begin{equation}
C \approx - 0.023 - 0.005\Omega^2
\end{equation}
$\Omega$ is the ratio of the spin period of the supergiant to its orbital period. We have assumed that the supergiant is close to Roche lobe-filling and will likely be rotating synchronously with the orbit, so $\Omega =1$, and the eclipse half angle is taken to be $\theta_{\rm e} = 34.9^{\circ} \pm 4.6^{\circ}$ \citep{hill05}.

Equations (2) - (10) allow the masses of the two stars to be determined in two limits. First, assuming that the system is viewed edge-on (in which case $i=90^{\circ}$) we can find a lower limit to the Roche lobe filling factor $\beta$ and lower limits to the stellar masses. Secondly, assuming that the supergiant fills its Roche lobe (in which case $\beta=1$) we can find a lower limit to the system inclination $i$ and upper limits to the stellar masses. The results in each limit are shown in Table 2. The true values of the masses of the component stars within the system may lie anywhere between these limits for corresponding combinations of $\beta$ and $i$. In order to determine the uncertainties on each derived parameter, we performed a Monte Carlo analysis which propagated the uncertainties through 10$^{4}$ trials, assuming a Guassian distribution in each individual uncertainty.  \\
Between phases $\phi$~ $\sim$~ 0.4 - 0.6 the He {\sc i} 1.70 $\mu$m absorption line displays asymmetric features for a number of the spectra (Fig 1). To investigate the effect these asymmetries have on the final orbital solution, we omitted spectra between this phase range and calculated a new set of NS and donor star masses. We found the impact on the derived orbital solution of these asymmetries to be negligible and have included the full set of measured radial velocities for completeness. We are unsure of the exact physical cause of these asymmetric features, but there appears to be a correlation between the airmass of the observation and the degree of asymmetry observed. We favour atmospheric effects arising at the time of the observation to be the cause; whilst acknowledging that they have a negligible effect upon the calculated system masses.

\section{IGR J18027-2016 Spectral classification} 
The H band spectrum is displayed in Fig. 3(a). Here we see the He {\sc i} 1.700 $\mu$m line is present but there is no detection of He {\sc ii} 1.692 $\mu$m, which is seen in absorption in early and late O supergiants and typically absent in early B supergiants. Further spectral indicators include the two hydrogen lines, Brackett 10 (1.736 $\mu$m) and Brackett 11 (1.681 $\mu$m) which display relatively weak absorption features indicative of B0 - B1 I \citep{hanson05}. \\
Examining the K$_{s}$ band, the absence of lines due to the C {\sc iv} triplet (2.069, 2.078 and 2.083 $\mu$m) combined with the He {\sc i} 2.058 $\mu$m line in emission implies IGR J18027-2016 is not an O type supergiant \citep{mason09}. Further constraining the spectral type, we can see that He {\sc i} 2.058 $\mu$m is in emission and when combined with the observed He {\sc i} 2.112 $\mu$m line in absorption, this is indicative of a B0-B2 I classification. The N {\sc iii} 2.115 $\mu$m line seen in emission allows us to again converge on a spectral type of B0-B1 I \citep{hanson05}. \\
We can also see that Br $\gamma$ displays relatively weak emission, unfortunately we have no spectral coverage of the He {\sc ii} 2.189 $\mu$m line region. As the spectral features in the K$_{s}$ band are not highly dependent on luminosity, we have difficulty distinguishing between luminosity sub-classes. We can not localise the classification to any better than B0 - B1 I. This is broadly in agreement with the spectral type found by \citet{torrejon10} of B1 Ib. IGR J18027-2016 does not appear to exhibit any secular changes in spectral morphology, both the K band spectrum obtained by \citet{torrejon10} and the spectrum shown in Fig. 3(b) are separated by 5 years but are very similar.

\section{Conclusions}
Within this paper we have presented the analysis and results of multi-epoch observations of the eclipsing high mass X-ray binary \object{IGR J18027-2016} performed at the ESO/VLT with the NIR spectrograph ISAAC. These have enabled us to make the first measurements of the dynamical masses of both system components. Constructing an orbital solution from near-IR radial velocity measurements and the orbital parameters of the system, provides a dynamically determined neutron star mass of 1.4 $\pm$ 0.2 - 1.6 $\pm$ 0.3 M$_{\odot}$. \\
The mass and radius of the supergiant donor, M $\sim$ 18 - 22 M$_{\odot}$ and R $\sim$ 17 - 20 R$_{\odot}$ is in agreement with that suggested from X-ray observations (M $\sim$ 21 M$_{\odot}$, R $\sim$ 19 R$_{\odot}$; \citet{hill05}).   \\ 
We find the spectral type of the donor to be B0-B1 I, this is in accord with that found previously (B1 Ib) by \citet{torrejon10}. Although we note that due to the lack of spectral coverage encompassing the He {\sc ii} 2.189 $\mu$m line, this prevented us from conducting non-LTE stellar atmosphere modelling and extracting the full range of stellar parameters. Particularly useful would have been a determination of the stellar temperature that would enable us to accurately constrain the luminosity sub-class. From an examination of a sample of Galactic B supergiants \citep{searle08} the donor has a wide possible range of luminosities, log(L/L$_{\odot}$) $\sim$ 5.5 - 5.7 and temperatures log T$_{\it eff}$ $\sim$ 4.3 - 4.5. Comparing this parameter range to evolutionary tracks (Fig. 8 \citet{searle08}) leads us to suspect that the donor's progenitor mass was $\sim$ 30 M$_{\odot}$.

\begin{acknowledgements}
ABM acknowledges support from an STFC studentship. JSC acknowledges support from an RCUK fellowship.
This research is partially supported by grants AYA2008-06166-C03-03 and
Consolider-GTC CSD-2006-00070 from the Spanish Ministerio de Ciencia e
Innovaci\'on (MICINN).
Based on observations carried out at the European Southern Observatory, Chile through programme ID 085.D-0539(A).
\end{acknowledgements}

\bibliographystyle{aa}
\bibliography{fifth_paper}

\end{document}